\documentclass[twocolumn,english,prl,showpacs,aps]{revtex4-1}
\usepackage[T1]{fontenc}
\usepackage[latin9]{inputenc}
\usepackage{amssymb}
\usepackage{times}
\usepackage{graphicx}
\usepackage{amsmath}
\usepackage{pstricks}
\usepackage{color}

\begin{document}
\title{Spontaneous solitons in the thermal equilibrium of a quasi-one-dimensional Bose gas}

\author{Tomasz Karpiuk,$^{1,2}$ Piotr Deuar,$^3$ Przemys{\l}aw Bienias,$^{4}$ Emilia Witkowska,$^3$ \\
Krzysztof Paw{\l}owski,$^{4,5}$ Mariusz Gajda,$^3$ Kazimierz Rz\k a\.zewski,$^{4,5}$ and Miros{\l}aw Brewczyk$^1$}
\affiliation{\mbox{$^1$Wydzia{\l} Fizyki, Uniwersytet w Bia{\l}ymstoku, 
                       ul. Lipowa 41, 15-424 Bia{\l}ystok, Poland}  \\
\mbox{$^2$Centre for Quantum Technologies, National University of Singapore, 3 Science Drive 2, Singapore 117543, Singapore }  \\
\mbox{$^3$Institute of Physics PAN, Al. Lotnik\'ow 32/46, 02-668 Warsaw, Poland}  \\
\mbox{$^4$Center for Theoretical Physics PAN, Al. Lotnik\'ow 32/46, 02-668 Warsaw, Poland}  \\
\mbox{$^5$Physikalisches Institut, Universit\"at Stuttgart, Pfaffenwaldring 57, 70550 Stuttgart, Germany} }

\begin{abstract}
We show that solitons occur generically in the thermal equilibrium state of a weakly-interacting elongated Bose gas, without the need for external forcing or perturbations. 
This reveals a major new quality to the experimentally widespread quasicondensate state, usually thought of as primarily phase-fluctuating. Thermal solitons are seen in uniform 1D, trapped 1D, and elongated 3D gases, appearing as 
shallow solitons at low quasicondensate temperatures, becoming widespread and deep as temperature rises.  
This behaviour can be understood via thermal occupation of the Type II excitations in the Lieb-Liniger model of a uniform 1D gas. Furthermore, we find that the quasicondensate phase includes very appreciable density fluctuations, while leaving phase fluctuations largely unaltered from the standard picture derived from a density-fluctuation-free treatment. 
\end{abstract}

\maketitle

Solitons, or non-destructible local disturbances, are important features of many one-dimensional (1D) nonlinear wave phenomena. In ultra-cold gases, they have long been sought, and were first observed to be generated by phase-imprinting\cite{Sengstock, Phillips}. More recently, their spontaneous formation in 1D gases was predicted as a result of the Kibble-Zurek mechanism\cite{ZurekI, Damski}, rapid evaporative cooling \cite{WitkowskaIII}, and dynamical processes after a quantum quench\cite{Gasenzer}. Here we show that they actually occur generically in the thermal equilibrium state of a weakly-interacting elongated Bose gas, without the need for external forcing or perturbations. 
This reveals a major new quality to the experimentally widespread quasicondensate state.
It can be understood via thermal occupation of the famous and somewhat elusive Type II excitations in the Lieb-Liniger model of a uniform 1D gas \cite{LiebII}.


A mathematically distinct class of soliton equations are the completely integrable systems. Among them, the Gross-Pitaevskii\cite{Pitaevskii, Gross} equation  describes weakly interacting bosons in a 1D geometry in the mean field approximation. The corresponding multi-atom Lieb-Liniger model of $N$ bosons on the circumference of a circle interacting by contact forces\cite{LiebI} has elementary excitations of two kinds: those of a Bogoliubov type and an additional ``Type II'' branch \cite{LiebII}. 
These additional excitations have been associated with solitons of the mean field\cite{Kulish, Kolomeisky, Jackson,Carr}. 
Although the trapping potential removes integrability, from the early days experimenters have searched for gray solitons. See \cite{Frantzeskakis} for a review. 

Typically, by irradiating one part of the condensate, one engineers a phase difference with the remainder, and a dark soliton forms at the interface between the phase domains\cite{Sengstock, Phillips}. 
Other proposed schemes involve taking the system away from equilibrium\cite{ZurekI, Damski, WitkowskaIII, Gasenzer}. 
Our results show that solitons are in fact present spontaneously even in equilibrium. However, engineered solitons have been easier to identify with the standard destructive imaging measurements because their position does not vary from shot to shot. 

Here we generate a classical field ensemble that describes the weakly interacting Bose gas at thermal equilibrium \cite{WitkowskaI}, and then show that gray solitons indeed are already there. This is demonstrated by tracing the time evolution of a single copy of the system. 
We find spectral properties consistent with Lieb Type II excitations. Moreover, we will also show that the presence of gray solitons in thermal equilibrium remains valid for very elongated traps that are no longer strictly 1D. 

Firstly, let us consider the simplest situation: a uniform 1D weakly-repulsively-interacting gas in free space with periodic boundary conditions, as per Lieb and Liniger \cite{LiebI}. To obtain the equilibrium state, we apply the classical field approximation (CFA) \cite{review,cfield}. Within this approach the usual bosonic field operator $\hat {\Psi }({z})$ which annihilates an atom at point ${z}$ is replaced by an ensemble of complex wave functions $\Psi ({z})$ obtained using a Monte Carlo sampling algorithm\cite{WitkowskaI} (see Supplementary material for details). 

In Fig. \ref{f1} we show the time evolution of the density and phase for single realizations in equilibrium at several temperatures. The starting conditions are randomly chosen from the collection of wavefunctions for the canonical ensemble obtained in the CFA. Subsequent time evolution of these chosen realizations in the CFA is via the Gross-Pitaevskii equation of motion.

The only relevant dimensionless parameter appearing in the Lieb-Liniger model of a 1D interacting Bose gas is $\gamma=mg/\hbar^2 \rho$, where $g$ characterizes the atom--atom interaction and $\rho=N/L$ is the linear density of the system. In what follows we will use $L$, $mL^2/\hbar$, $\hbar^2/mL^2$, and $\hbar^2/mL^2 k_B$ as the units of length, time, energy, and temperature, respectively.

\begin{figure}[thb]
\includegraphics[width=4.2cm]{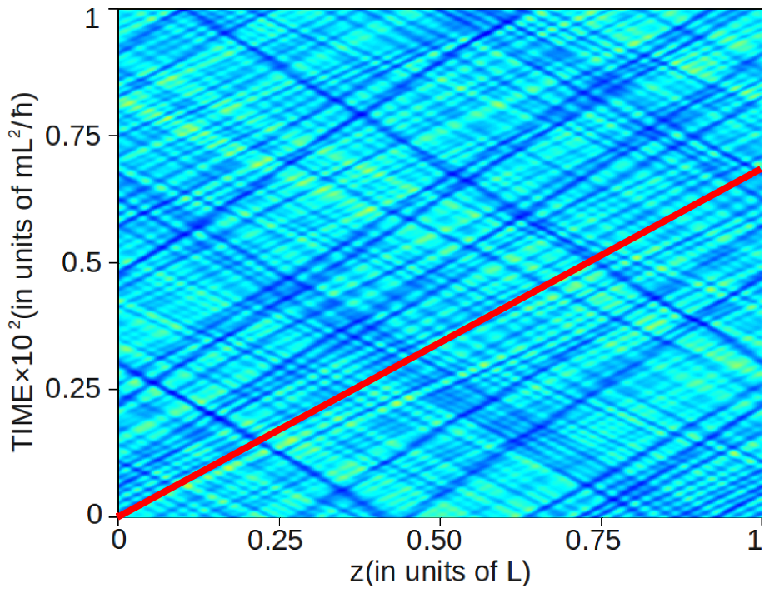}
\includegraphics[width=4.2cm]{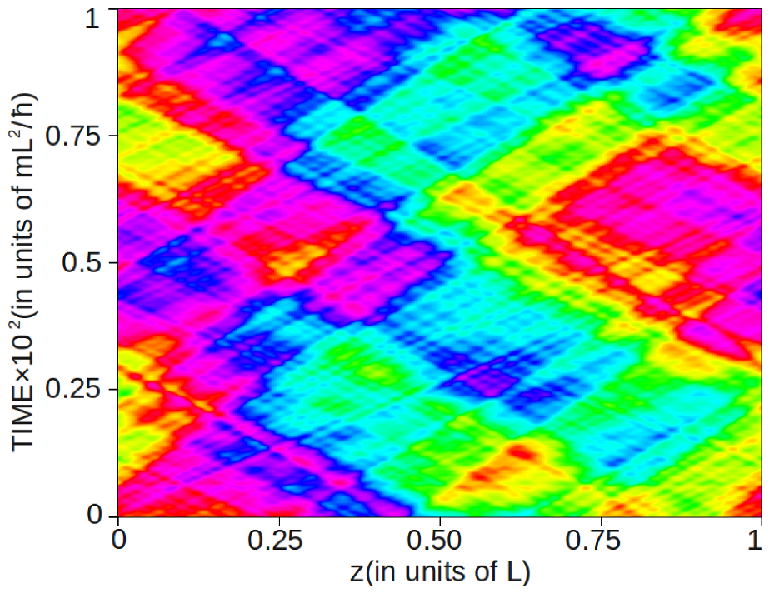}\\
\includegraphics[width=4.2cm]{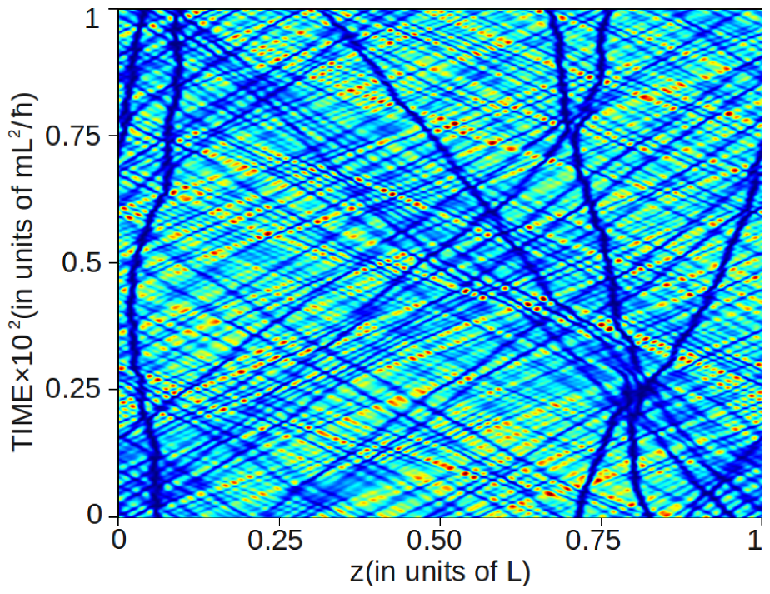}
\includegraphics[width=4.2cm]{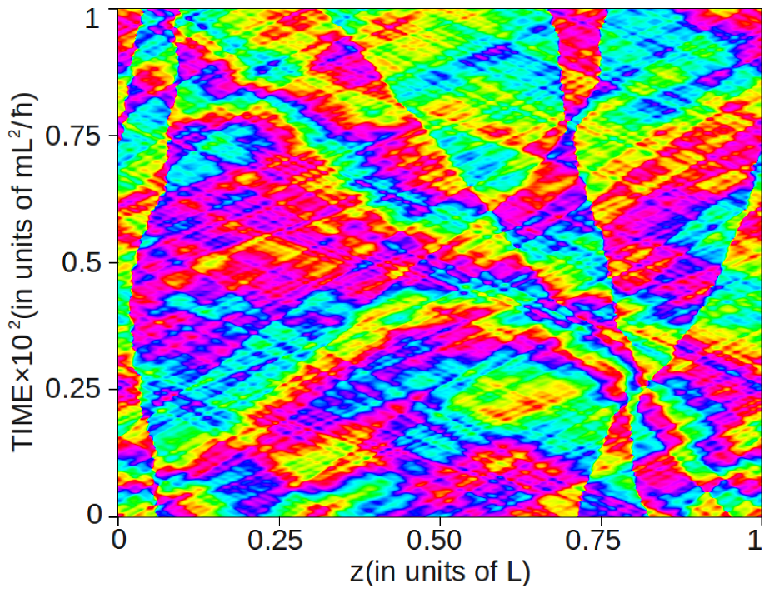}\\
\includegraphics[width=4.2cm]{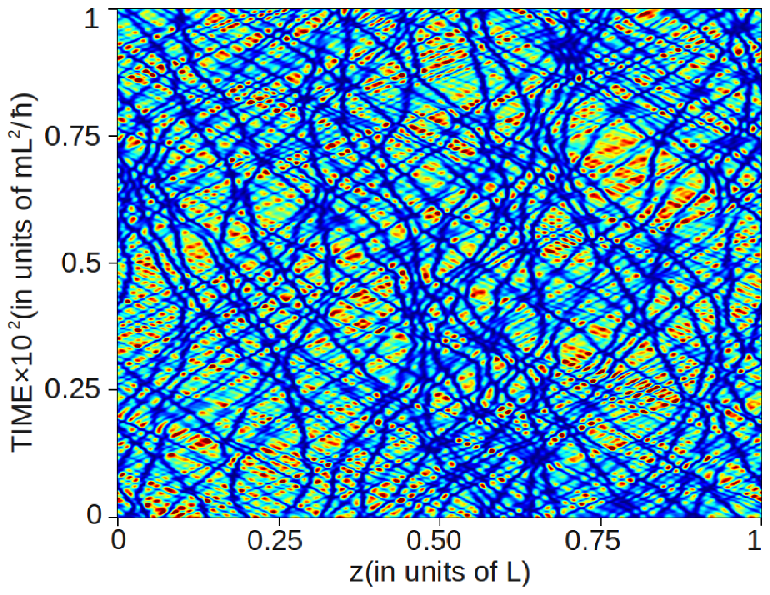}
\includegraphics[width=4.2cm]{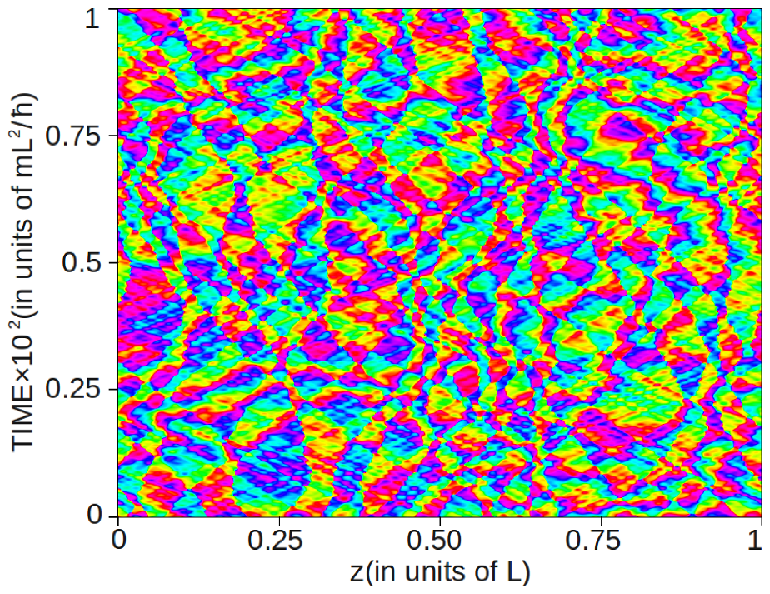}
\caption{
Density (left) and phase (right) as a function of time of a 1D Bose gas at equilibrium for a single realization. Here $\gamma = 0.02$ ($N=10^3$) and the temperatures are $T=10^4=\mu/2$ (top), $7\times 10^4$ (middle), and $10^5$ (bottom) in units of $\hbar^2/mL^2 k_B$. The red line in the upper left panel corresponds to travel at the speed of sound.
}
\label{f1}
\end{figure}

Firstly, the top panels show the system at the relatively low $T=10^4=\mu/2$. 
There are numerous density disturbances in the form of dips and peaks, traveling in both directions, near the speed of sound (which is depicted by the red line). There exist both defects that travel faster, and those that travel slower than sound. The deepest dips are slower -- the slope of their trajectories on the plot is higher than that of the red line. 
At low temperature only fast moving shallow defects are present. Some of them correspond to packets of Bogoliubov phonons, others are solitons. This can be checked by inspection of the phase jump across the defect (for a soliton its sign is related to its direction of movement), and by fitting the local density dip to a soliton solution\cite{Zakharov}. However, shallow fast solitons can be hard to tell from phonons.

For a higher temperature $T=7\times10^4$ (middle panels), long-lived deep dips travelling far slower than the speed of sound appear. These become prolific as the temperature increases to $T=10\times10^4$ (lower panel). An analysis of the parameters of slower-than-sound defects confirms their interpretation as dark solitons. Indeed, one sees the anti-correlation between the depths and speeds of the dips that is expected for dark solitons. Near maximal depth the dips approach being stationary. One also sees numerous soliton collisions, and associated phase-shifts, with the number of solitons usually conserved.  

\begin{figure}[thb] 
{
\includegraphics[width=7cm]{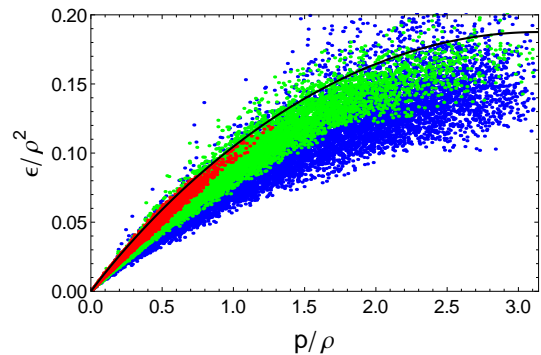}
}
\caption{Comparison of the spectra of Lieb Type II excitations (solid black line) and dips in the thermal state for three different temperatures: $T=10^4$ (red dots), $4\times 10^4$ (green dots), and $10^5$ (blue dots). Here $\gamma = 0.02$ ($N=10^3$).
\label{liebsol}}
\end{figure}

The  match between dark solitons and Lieb Type II excitations is seen by comparing fitted parameters describing the local density dips to the grey soliton solutions. 
To do this, we fit a density profile (characterised by soliton depth, local density and soliton position) to each dip. If the rms parameter errors are smaller than 10\% of the local density (for depth or density) or the healing length (for position) we accept the dip as a soliton. We then calculate soliton energy $\epsilon$ and momentum $\tt p$ from the parameters\cite{Kulish}, collecting data from many time snapshots. The resulting energy-momentum relation of the accepted dips is compared with the Lieb Type II excitation spectrum in Fig. \ref{liebsol} for three temperatures. 
At the lowest temperature, the spectral match is good, and includes the negative curvature of the type II branch. 
For higher temperatures, the fitted soliton energy lies somewhat below the Lieb II line, retaining a qualitative match. Interpretation of the shift is difficult as interactions between solitons are frequent at this temperature. Notably, one sees a filling of the high $\tt p$ spectrum with temperature.

This allows us to explain the temperature dependence of the appearance of deep solitons. 
These have the maximum Lieb II energy and momentum\cite{Jackson}. For the parameters of Fig.~\ref{f1}, that  is about $10^5$ in units of temperature (see Supplementary material). Thus we would expect the deepest solitons to appear when temperatures of this order are reached. This is what is seen. We found similar agreement for $\gamma=0.002$. 

It is appealing to think of the  
system as of a gas of quasiparticles of two kinds -- where  \emph{two} bosonic excitation families coexist like in the Lieb-Liniger model (see Supplementary material). These two kinds -- phonons and solitons, interact with each other in an inelastic way, so that solitons can be born from Bogoliubov excitations and vice versa. Such interactions could allow the gas of quasiparticles to reach the equilibrium state. 

\begin{figure}[thb]
{
\mbox{}\hspace*{-0.2cm}\raisebox{3cm}{\begin{minipage}{4cm}\includegraphics[height=4cm]{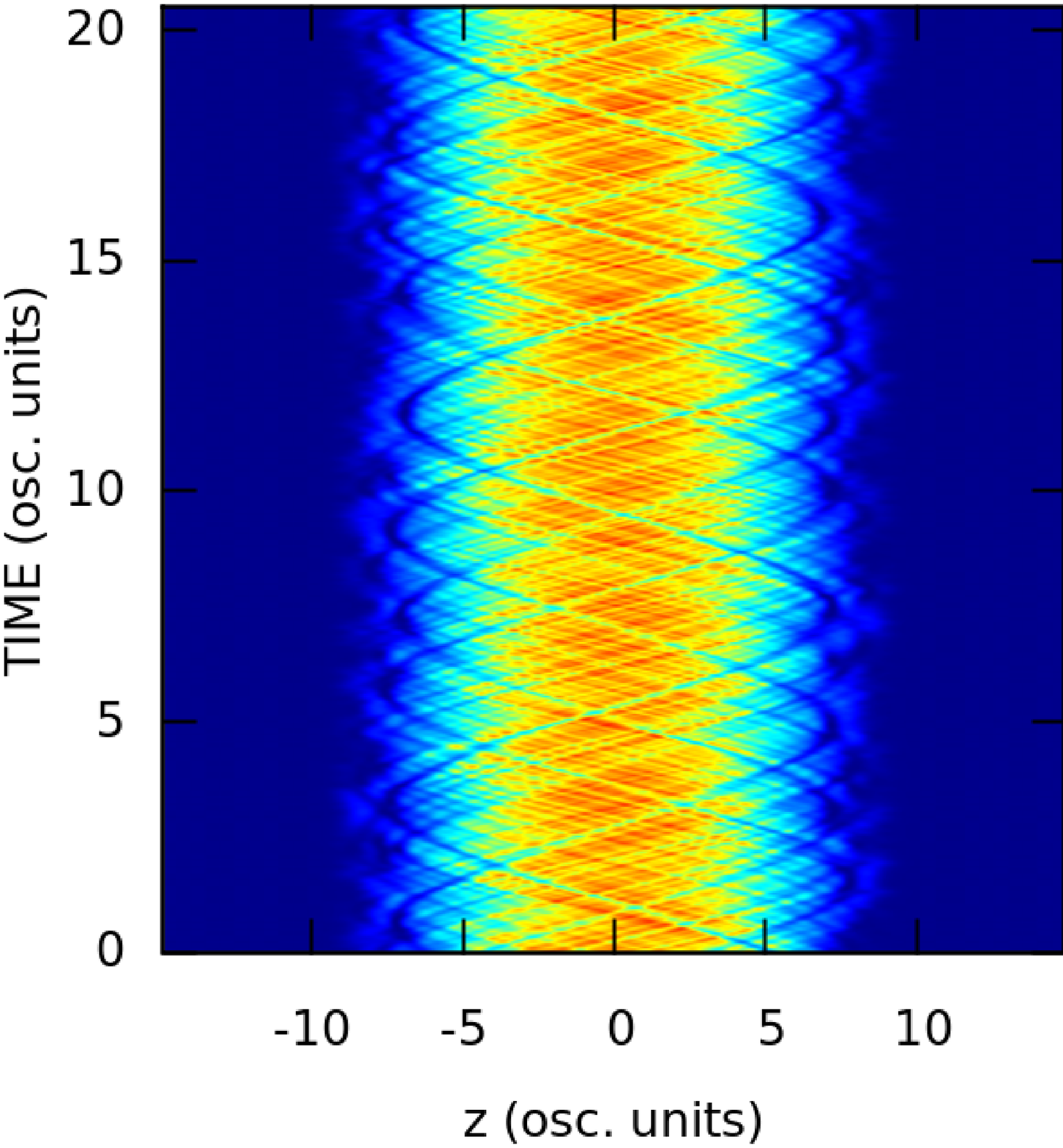}\\\includegraphics[height=4cm]{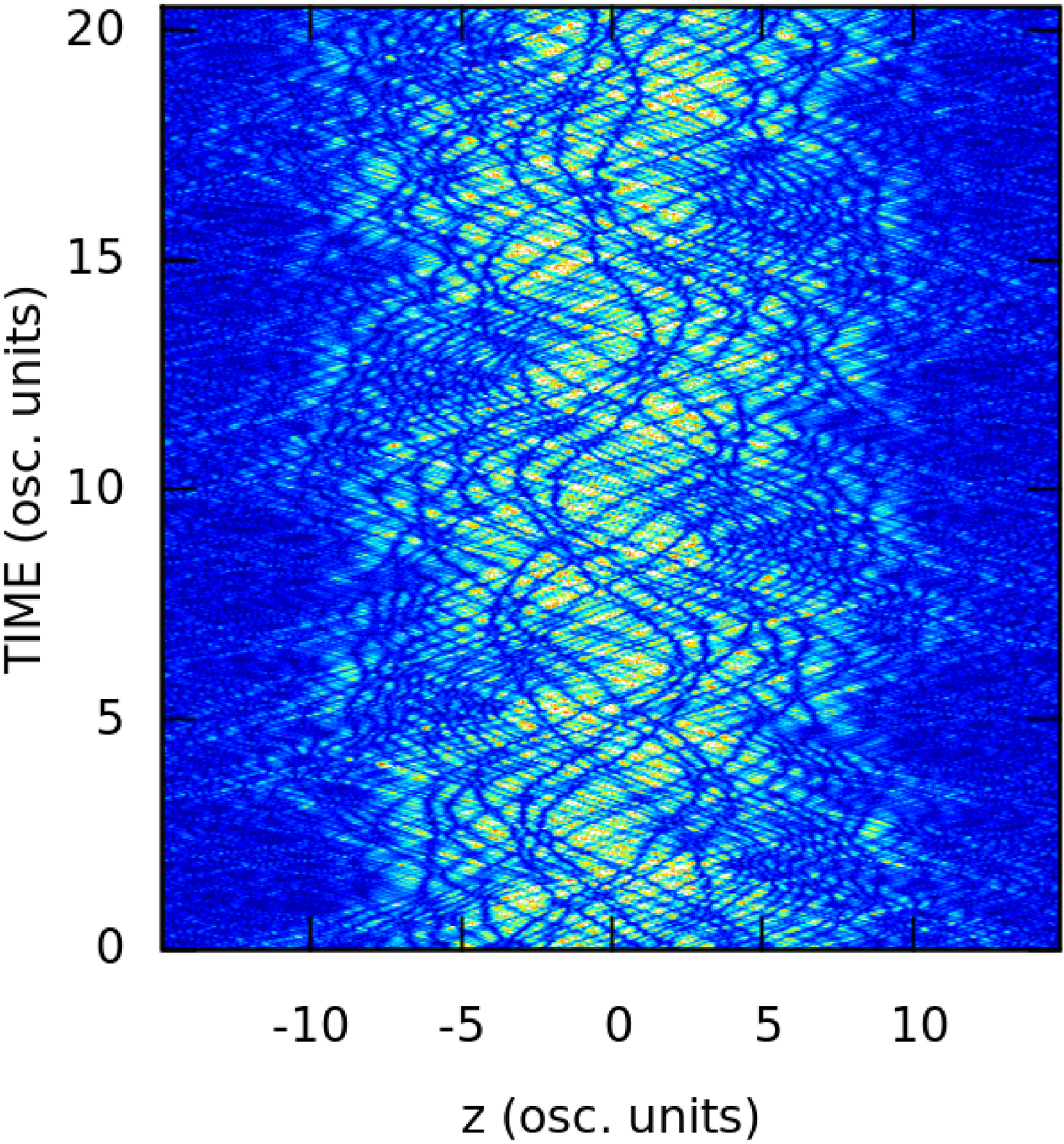}\end{minipage}}
\raisebox{3cm}{\begin{minipage}{4cm}\includegraphics[height=8cm]{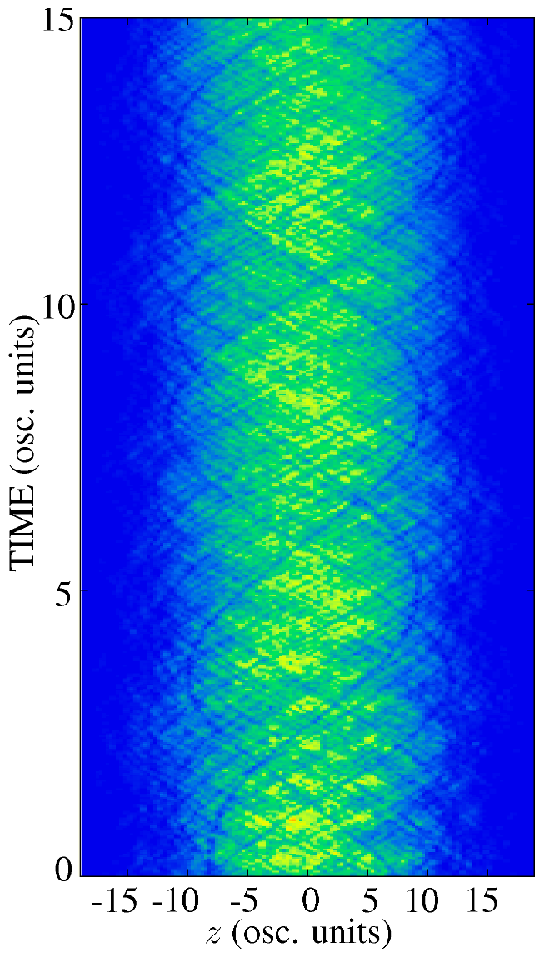}\end{minipage}}
\caption{Solitons in trapped clouds. The left panels show the time-evolution of a trapped 1D condensate with parameters: $N=1000$, 1D interaction strength $g=0.31$ in units of $\hbar \omega_z (\hbar / m \omega_z)^{1/2}$, and trap frequency $\omega_z=2\pi\times 10\,$Hz. The temperatures are: $k_BT=15\hbar\omega_z$ (upper frame) and $k_BT=260\hbar\omega_z$ (lower frame). The right panel shows the evolution of the central density in a three-dimensional elongated trapped cloud (horizontal direction). Here, the trap frequencies are: $\omega_z=2\pi\times 10\,$Hz and  $\omega_{\bot}=2\pi\times 1000\,$Hz, with again $N=1000$, and the (3D) interaction strength $g_{3D}=0.213$ in units of $\hbar \omega_z (\hbar / m \omega_z)^{3/2}$. The temperature is $80\,\hbar \omega_z/k_B$.   } 
\label{f4}}
\end{figure}

Now a question that begs to be asked is whether solitons will also be present in the less idealized case of a trapped gas in the quasicondensate regime -- a situation ubiquitous in contemporary experiments. The CFA can be applied in a similar way as above (see Supplementary material). Firstly, consider a harmonically-trapped 1D gas. Fig. \ref{f4} (left) shows its  density evolution for two different temperatures. The lower temperature case (left upper panel) has $T$  chosen such that the phase coherence length $l_{\phi}$ is approximately equal to the size of the cloud $W$, i.e. it corresponds to the temperature  $T_{\phi}$ which separates the quasicondensate and true condensate phases, as first described by Petrov \textit{et al.}\cite{Petrov}. 
For temperatures lower than this, only shallow fast solitons are seen in the bulk of the cloud.
 At much higher temperature (lower left panel in Fig. \ref{f4}) a mass of deeper long-lived solitons emerges. They remain within the main part of the atomic cloud, and  their number and depth increases with temperature. One can easily check that for the lower temperatures the solitons oscillate in the trap with a
frequency which within a few percent agrees with  $ \omega / \sqrt{2}$ from a simple model\cite{Busch}. At high
temperatures, inter-soliton collisions come to dominate the oscillating behavior, making that model inappropriate.

\begin{figure}[hbt]
\includegraphics[width=6cm]{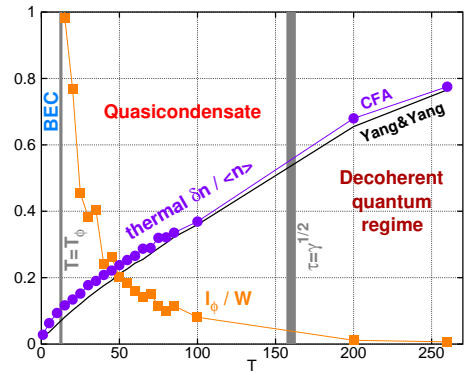}
\caption{Phase and thermal density fluctuations in the center of the trapped 1D gas as a function of temperature. Parameters are as in Fig.~\ref{f4} except for varying $T$ (in units of $\hbar\omega_z/k_B$). Purple circles show \mbox{$\delta n/\langle n\rangle$} in the center of the trap; Orange squares the ratio of the phase coherence length $l_{\phi}$ to the cloud width $W$. The latter is taken to be the distance between points at which the density has fallen to 10\% of the central value. Grey bars indicate temperatures $T_{\phi}$  and an estimate of the location of the quasicondensate / decoherent quantum gas crossover at $\tau=\sqrt{\gamma}$ \cite{Kheruntsyan03}. Yang\&Yang predictions (black solid line) are for the average density in the center ($|z|\le2$) of the cloud. Some statistical uncertainty is visible.
\label{figdn}}
\end{figure}

The center of the cloud can be considered as a uniform gas under a local density approximation (LDA) as confirmed in recent experiments\cite{vanAmerongen08,Jacqmin11}. 
This is described by the dimensionless interaction strength ($\gamma=g/\langle n\rangle$) and temperature ($\tau=2k_BTm/\hbar^2\langle n\rangle^2$). In Figures~\ref{f4} and~\ref{figdn}, the $T=15$ gas lies deep in the quasicondensate regime ($\gamma=0.0033, \tau=0.0034$), while the center of the $T=260$ cloud is a decoherent quantum gas\cite{Kheruntsyan03} ($\gamma=0.0059, \tau=0.19$). Here, Fig~\ref{figdn} shows that significant local density fluctuations $\delta n/\langle n \rangle \approx 10-40\%$ are present in the quasicondensate phase $\tau\ll\sqrt{\gamma}$. This is to be contrasted with the common view that the only notable fluctuations in the quasicondensate are those of the phase, a situation we see only at the very lowest temperatures. (Note also the long-wavelength density fluctuations in\cite{Jacqmin11}). 

Importantly, the CFA fluctuations agree with predictions obtained from the exact local density correlation function\cite{Kheruntsyan03} in the Yang \& Yang description\cite{YangYang} of the interacting uniform gas. Note that we have taken account of the fact that the CFA does not include zero temperature fluctuations, so that the $\delta n$ in Fig.~\ref{figdn} shows only the thermal contribution. The equivalent contribution in the Yang and Yang solution is $(\delta n)_{\rm thermal}/\langle n\rangle = \sqrt{g^{(2)}(0) - g^{(2)}_{T=0}(0)}$, which is shown as a black line. The level of agreement implies that the solitons we have seen here in Fig.~\ref{f4}, that contribute a large proportion of the density fluctuations, must also be an essential feature of the Yang\&Yang description. 

We have also calculated the coherence length through matching the phase correlation function $g^{(1)}(z,z')=\rho_1 (z,z')/\sqrt{\rho_1 (z,z) \rho_1 (z',z')}$ to an exponential decay $\exp\left[-|z-z'|/l_{\phi}\right]$. This is shown in Fig.~\ref{figdn} in orange, and is still in agreement with the canonical treatment that ignores density fluctuations\cite{Petrov}. The one-particle density matrix, $\rho_1 (z,z')$, is obtained by an averaging of \mbox{$\langle\Psi^* (z) \Psi(z')\rangle$} over the initial canonical ensemble.  For the low temperature case ($T=15$), $l_{\phi} = 13.7a_{\rm ho} = 13.7\sqrt{\hbar/\omega_zm}$, while the cloud width $W$ is $14.7a_{\rm ho}$. To compare, the standard quasicondensate expressions in \cite{Petrov} that consider only phase fluctuations give an estimate of the temperature at which $l_{\phi}=W$. It is
$T_{\phi}^{\rm phase\ only}  = \frac{3N(\hbar\omega_z)^2}{8\mu k_B} = 12.5\,\hbar\omega_z/k_B$ for these parameters, which is a good agreement with the CFA calculation. 
Our agreement on $T_{\phi}$ with the standard description is so good because it is a manifestation of the fact that the correlation length obtained from a pure phase dependence of the classical field, i. e. $g^{(\phi)}(z,z')=\exp[i(\phi(z)-\phi(z^\prime))]$ like in \cite{Petrov} --- where $\phi(z)$ is the phase at point $z$ --- does not differ significantly from what is obtained via the complete correlation function $g^{(1)}(z,z')$\ \cite{Bienias}. We conclude then, that the 10-40\% density fluctuations and solitons that we see in the quasicondensate, are not inconsistent with the past calculations of phase fluctuations. \emph{Rather,} they are an inherent feature that had remained un-noticed until this time due to the prevalent focus on phase coherence.

These results raise an interesting issue regarding the Kibble-Zurek mechanism of defect formation (KZM)  in 1D gases. 
In a recent paper \cite{ZurekI} it was conjectured that dark solitons should be created by the KZM during rapid cooling of the gas\cite{Kibble,Zurek}. 
Simulations\cite{WitkowskaIII} of the cooling of a 1D Bose gas in a harmonic trap indeed revealed the presence of solitons. However, we know now that some solitons are present even in thermal equilibrium. Only in the case when the number of solitons at the end of the cooling process exceeds that found in its corresponding thermal equilibrium,  could simulations of that kind confirm the KZM at work.

Finally, there is a question whether these solitons can still survive when the strict 1D trapped system crosses over to one in a very elongated cigar-shaped trap. 
This is made plausible by past work which showed that quasicondensate-BEC-like transitions occur for cigar-shaped systems analogously with the true 1D gas\cite{Petrov1, Kadio,qcexp}. 
The right panel of Fig.~\ref{f4} shows the evolution of a thermal state in a fully three-dimensional calculation. Solitons are still clearly visible. This indicates that they may be present, and indeed even widespread, in many existing experiments. 

\begin{figure}[thb]
{
\includegraphics[width=7cm]{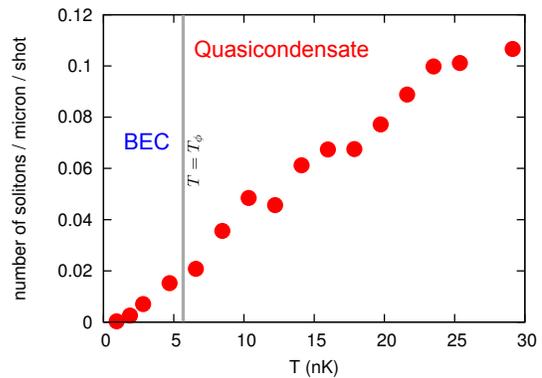}
}
\caption{
Number of solitons per $\mu$m on the edges of the cloud, per snapshot, versus temperature. There are  $10^4$ Rb$^{87}$ atoms in a 1D trap with $\omega_z = 2\pi \times  1.9$Hz.
The 1D coupling assumes a Gaussian transverse profile corresponding to $\omega_{\perp} = 2\pi \times 128$Hz.
Counting was over 7$\mu$m intervals on either side of the cloud centered around the
position where the average density is $10\mu$m$^{-1}$ (12\% of peak). Dips whose central density falls to below $10$\% of the mean density at their trap position were counted.
\label{solnum}
}
\end{figure}

The observation of thermal solitons with existing equipment has been non trivial because of two factors: The soliton width in-situ was typically significantly narrower than the detector resolution, and an identification of solitons  by ``eyeballing'' is only straightforward if one has access to observations of the in-situ dynamics, not destructive snapshots. 
Looking at the edges of the gas offers an indirect way of detecting the solitons\cite{Oberthaler}. As they bounce
between the edges of the gas, they approximately retain their absolute
depth while also increasing in width and reaching close to 100\% relative depth at the turning point, making them accessible with realistic resolutions. 
We do not see correspondingly deep phonon disturbances in our images. 
Therefore \cite{Oberthaler}, we
have counted the density of dips with over 90\% relative depth in the wings of the 1D cloud. The growth of the density of these solitons with
temperature is shown in Fig. \ref{solnum}.

Further progress could be made by recent experimental advances such as high-resolution detection based on scattering of an electron beam\cite{OttSM}, analysis of momentum spectra\cite{Gasenzer}, phase-jump statistics\cite{Schmiedmayer} (phase domains are larger than soliton dips) or with long\cite{Sengstock, Phillips, Sengstock1SM} or anti-trapped expansions\cite{ChangEngelsSM} (See Supplementary material for more detail).

In summary we have found that solitons are a natural and spontaneous feature of the 1D and elongated weakly-interacting Bose gases. While solitons have been studied in such systems before by imprinting, the above results show that they are in fact common. The good agreement with exact density fluctuations in Fig.~\ref{figdn} shows that quantitative study of these phenomena is possible with the CFA, and possibly related methods. The 1D Bose gas at thermal equilibrium is a system where two very different kinds of {bosonic} excitations are present simultaneously. The system actually contains much more than the standard picture, especially in the quasicondensate regime. Here, in addition to the well known phase fluctuations, appreciable density fluctuations are found, including deep solitons.
These spontaneous solitons in the thermal state are analogous to pairs of vortices present in a 2D
gas near the Berezinskii-Kosterlitz-Thouless transition \cite{Berezinskii}. 

\textsl{Acknowledgements:} 
We are grateful to Matthew Davis, Tilman Pfau, and Tomasz \'Swis{\l}ocki
for helpful discussions. T.K. acknowledges support by the National Science Center Grant No. 2011/01/B/ST2/05125. P.D. by the Polish Government research grant N N02 128539, and P.B., K.P., and K.R. by Polish Government research grant N N202 174239 both for the years 2010--2012. M.G., E.W., and M.B. acknowledge support from the EU NAMEQUAM project, P.B., K.P., and K.R. acknowledge financial support by contract research `Internationale Spitzenforschung II-2' of the Baden-W\"urttemberg Stiftung, ``Decoherence in long range interacting quantum systems and devices''. The CQT is a Research Centre of Excellence funded by the Ministry of Education and the National Research Foundation of Singapore.

\clearpage

\renewcommand{\figurename}{Figure S\!}
\setcounter{figure}{0}
\textcolor[rgb]{1,1,1}{
\begin{widetext}
\begin{center}
\textcolor[rgb]{0,0,0}{
\bf Supplementary material for: \\  Spontaneous solitons in the thermal equilibrium of a quasi-one-dimensional Bose gas
}
\end{center}
\vspace*{1cm}
\end{widetext}
}


\subsection{The classical field approximation}
Within this approach \cite{Sreview,Scfield} the usual bosonic field operator $\hat {\Psi }({z})$ which annihilates an atom at point ${z}$ is replaced by the complex wave function $\Psi ({z})$. Technically speaking, we first expand the field operator in the basis of one-particle wave functions, appropriate for the problem considered. Then, extending the original Bogoliubov idea \cite{SBogoliubov} to all macroscopically occupied one-particle modes, we replace the operators corresponding to these modes by $c$-numbers. Restricting the expansion only to these modes, the field operator is turned into a complex wave function -- the classical field. For a plane-wave basis,
\begin{equation}
\Psi ({z}) = \sum_{{| k|}\le{k_{\rm max}}} \alpha_{ k}\,  \frac{1}{\sqrt{L}}\, e^{i {k} {z}}  \,.  
\label{expansion}
\end{equation}
The above expression is appropriate for bosons confined in a box of length $L$ with periodic boundary conditions. For a harmonic trapping potential, the single-particle trap eigenfunctions are more convenient. The summation is extended over all modes up to the momentum cut-off $\hbar k_{\rm max}$. The optimal choice of the cut-off is discussed in Ref. \cite{SWitkowskaII}. Fully three dimensional, elongated trap simulations require recalculation of the optimal cut-off condition since the explicit results in \cite{SWitkowskaII} are valid only for symmetric D-dimensional traps and, moreover, are asymptotic for large number of atoms. Each classical field, $\Psi ({z})$, shares many properties with the single-shot measurements of the atomic cloud that occur in experiment. The classical field satisfies the following equation of motion \cite{Sreview}:
\begin{eqnarray}
&&i\hbar \frac{\partial}{\partial t} {\Psi }({{z}},t) =
\left( -\frac{\hbar^2}{2m} \frac{\partial^2}{\partial z^2} + g\, |{\Psi }({{z}},t)|^2  \right)
{\Psi }({{z}},t) \,,  
\label{CFequation}
\end{eqnarray}
where $g$ characterizes the atom--atom interaction and the nonlinear term is projected on the subspace spanned by the macroscopically populated modes. \\

\subsection{Thermal states}
To obtain the thermal equilibrium state of a 1D Bose gas within the CFA, we numerically generate members of the canonical ensemble of states \cite{SWitkowskaI}, i.e. states populated according to the probability distribution given by:
\begin{equation}
P(\{\alpha_{k}\}) = \frac{1}{Z} e^{-E_{\Psi} / k_B T} \,,
\label{probability}
\end{equation}
where $Z$ is the canonical partition function and $T$ is a temperature. The energy, $E_{\Psi}$, accumulated in the classical field is given by :
\begin{equation}
E_{\Psi} = \int_L d{z}   \Psi^* ({z})H_0 \Psi ({z}) +
\frac{1}{2} g \int_L d{z} |\Psi ({z})|^4  \,.
\label{energy}
\end{equation}
where $H_0$ is the single-particle Hamiltonian 
\begin{equation}
H_0 = V_{\rm trap}(z) -\frac{\hbar^2}{2m}\frac{\partial^2}{\partial z^2} 
\end{equation}
with the trapping potential $V_{\rm trap}(z)$.
An extra constraint on the amplitudes $\{\alpha_{k}\}$ should be fulfilled:
\begin{equation}
\sum_{{|k|}\le{k_{\rm max}}} |\alpha_{k}|^2 = N  \,,
\label{constraint}
\end{equation}
where $N$ is the number of atoms. An efficient way to obtain states belonging to the canonical ensemble at given temperature $T$ is a Monte Carlo method using the Metropolis algorithm\cite{SMetropolis}. Here, a random walk in the phase space of the system is performed and all visited states become the members of the canonical ensemble. These states are used to calculate statistical averages of any observable. Details for ultracold Bose gas systems are given in Ref.~\cite{SWitkowskaI}.

\subsection{Temperature at which deep solitons appear}
Assuming the excitations observed within the CFA are Lieb type II excitations, we are able to estimate the temperature at which deepest solitons appear in the thermal state of a uniform system. To obtain the dispersion curve for the type II excitations, one is required to solve the inhomogeneous Fredholm integral equations \cite{SLiebII}. This can be done numerically, and for $\gamma=0.02$ the dispersion curve, i.e. energy versus momentum $\epsilon(p)$, is found to be well approximated by $\epsilon(p)=a \rho p + b p^2$, where $a=0.125$ and $b=-0.021$. Maximal excitation energy occurs  for the highest momentum on the curve $p_{\rm max}=\pi N$ (see \cite{SLiebII} for details), and corresponds to the deepest, stationary solitons. It is then given by the formula $\epsilon_{\rm max}=\pi N^2 (a+\pi b)$ in our units, which for $N=1000$ is about $10^5$. Therefore, for $\gamma=0.02$, the deepest solitons are expected at temperatures of this order or higher.

\subsection{Spectral analysis}
\begin{figure}[thb] 
{
\includegraphics[width=5cm]{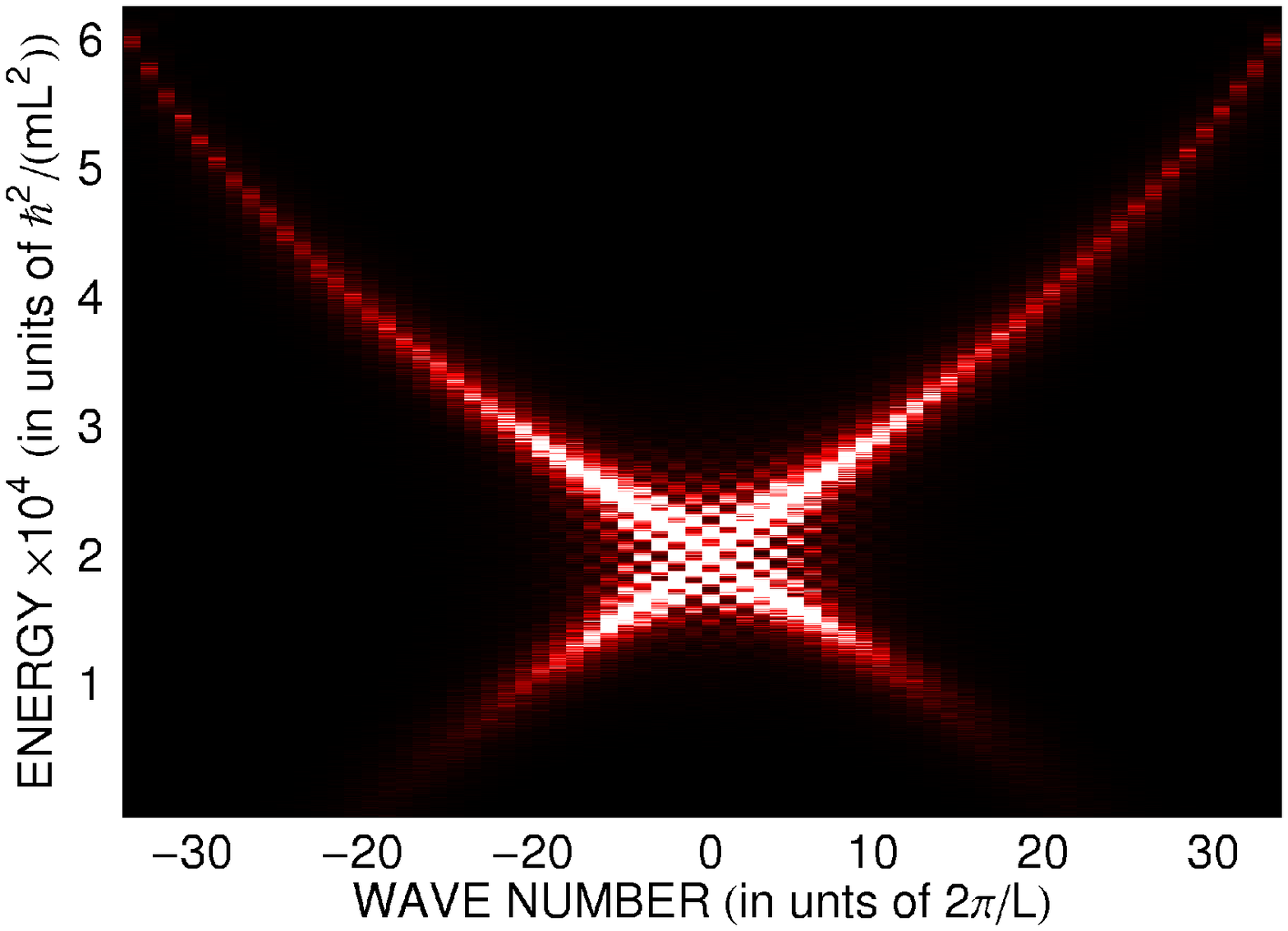}\\
\includegraphics[width=3cm]{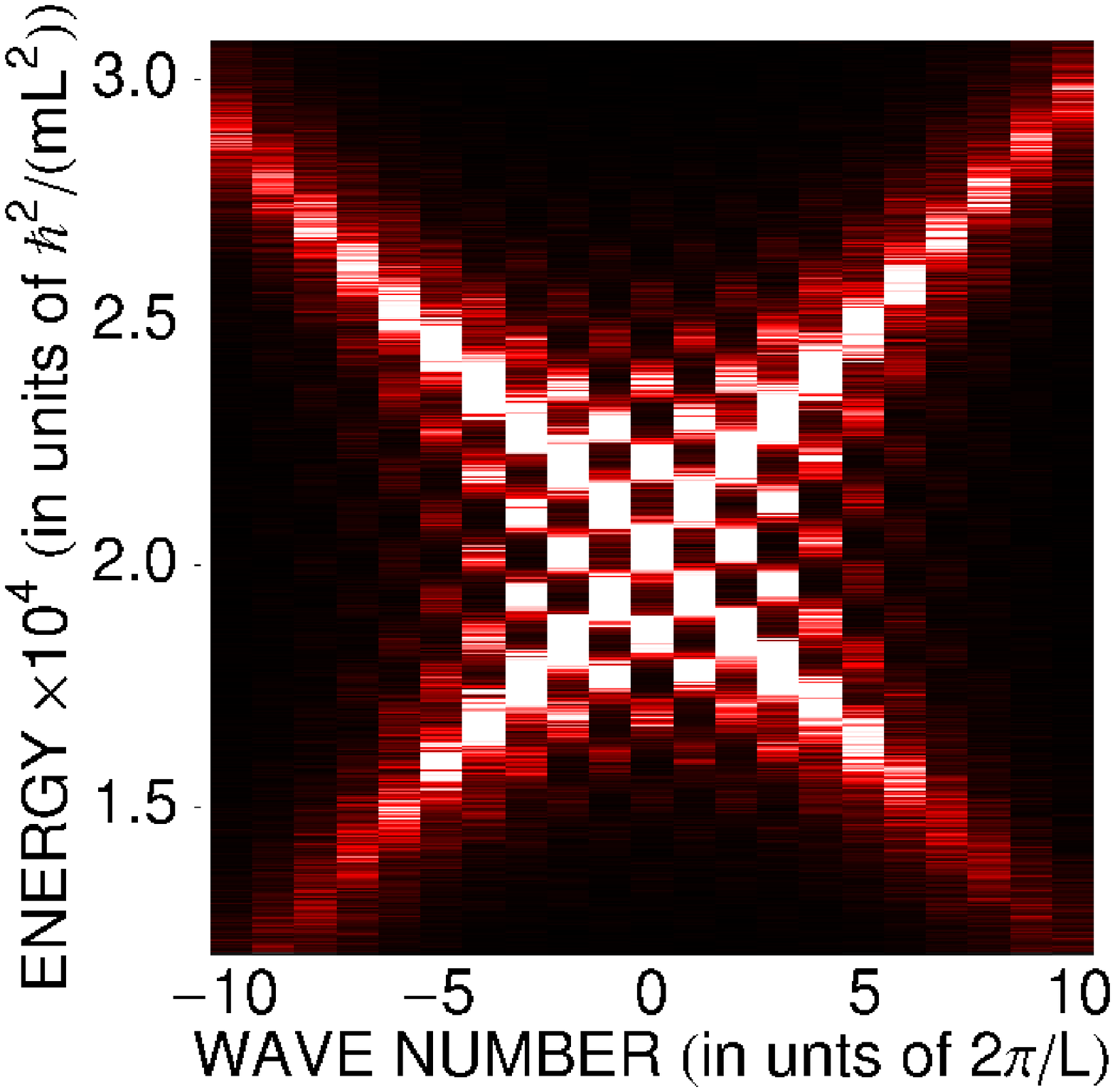} \includegraphics[width=3cm]{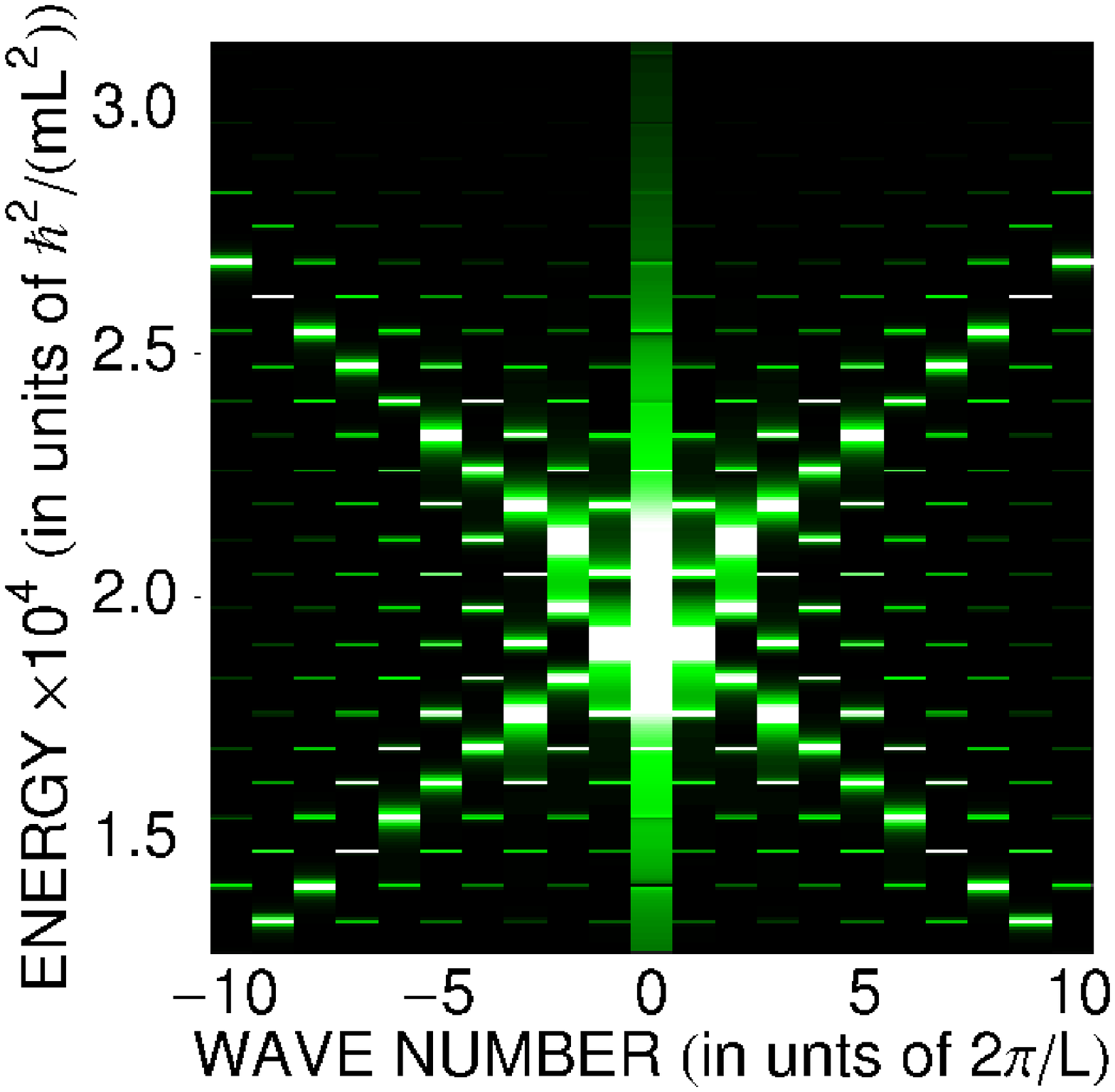}
}
\caption{Spectral density of a 1D Bose gas at equilibrium for $\gamma=0.02,T=10^4$  within the classical field approximation (upper panel, detail -- lower left panel). 
In addition to the Bogoliubov phonon nature of the main curves, the spectral density exhibits a checkerboard pattern for low momenta. This pattern is related to the presence of solitons in the system. Bottom right panel: spectral density of two dark solitons propagating in opposite directions. Soliton velocities are $\pm0.8 c$, with $c$ the speed of sound. The checkerboard pattern again appears on top of the main structure consisting of two lines $\omega=\pm uk$. It is interpreted as a result of interference between solitons.
\label{spectrum}}
\end{figure}

Further arguments for the simultaneous presence of both kinds of excitations (Type I -- Bogoliubov phonons, and Type II -- solitons) in a 1D Bose gas at the equilibrium come from a spectral analysis of the classical field. Fig. \ref{spectrum} shows the space-time spectral density $|\Psi(\omega,k)|^2$ of a single realization at equilibrium. Like in the three-dimensional case (see Ref. \cite{S3Dbox}), two curves crossing at a frequency equal to the chemical potential $\mu=2\times10^4$ are clearly visible. The low-momentum slope of each curve is just the speed of sound. The parts of these curves below the chemical potential are necessary to construct the phonon-like section of the excitation spectrum \cite{S3Dbox,SZawitkowski} and fade at momenta at which the dispersion curve changes its character from linear to parabolic. This part of the spectrum proves the existence of Bogoliubov phonons in thermal equilibrium.

However, in addition to the phonon-like behavior, the spectral density exhibits a phenomenon absent in the three-dimensional system -- in the region of low energies and momenta a checkerboard pattern appears (see Fig. \ref{spectrum}). This is a signature of the presence of solitons. We confirm this interpretation by considering the spectral density of
 two counter-propagating dark solitons moving with the same velocity $u$ (and depth), shown in the lower right panel of Fig.~\ref{spectrum}. These solitons are obtained from 
the Zakharov solution \cite{SZakharov}, and its subsequent evolution according to the nonlinear Schr\"odinger equation. Finally, the dispersion curve of a single dark soliton consists of only a single line $\omega=u k$. Hence, the checkerboard pattern appears as a result of interference between the solitons, and thus it is a strong signature of the existence of dark solitons in the system. Indeed, the checkerboard pattern becomes more regular when the number of solitons increases.

\subsection{The classical field approximation in one dimension}
  The classical field approach\cite{Sreview}  and the very closely related PGPE (Projected Gross-Pitaevskii Equation) approach\cite{Scfield}  have been benchmarked on numerous 
occasions --- see description in the reviews \cite{Sreview,Scfield}, or \cite{SKarpiuk-freq}. To demonstrate its applicability to the density fluctuations of 1D gases as considered in this article, we have compared CFA results to a recent experiment that measured density fluctuations\cite{SBouchoule}.

\begin{figure}[hbt]
\includegraphics[width=8.6cm]{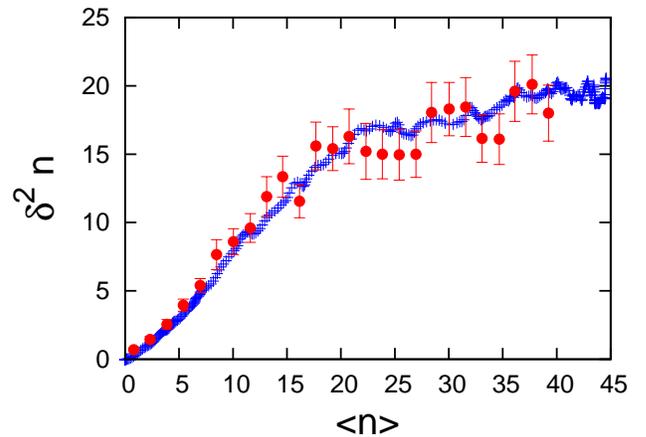}
\caption{Local atom-number variances, $\delta^2n$, as a function of the mean local atom density in a weakly interacting  quasi 1D gas confined by a harmonic potential.  Red circles are taken from Fig. 1c. in \cite{SBouchoule}. Blue crosses are the results of our 1D simulation with parameters from the experiment \cite{SBouchoule}: $g_{1D} = 2 \hbar \omega_{\perp} a = 2.06 \hbar\omega_z(\hbar/m\omega_z)^{1/2}$ is the coupling constant, $a=5.7$nm is the 3D s-wave scattering length, and $\omega_{\perp} = 2\pi\times 3.9$ kHz, $\omega_z = 2\pi\times4$ Hz are the frequencies of the transverse and longitudinal harmonic confining potentials, respectively. $T=0.09\hbar\omega_{\perp}/k_B = 88\hbar\omega_z/k_B$.  }
\label{fluctuations}
\end{figure}

Using the actual parameters of \cite{SBouchoule} and the prescription for the cut-off parameter used in this article, we obtained the very good agreement illustrated in Fig S\ref{fluctuations}.

\subsection{Possible detection schemes}
Let us now turn to the question of experimental observation of solitons in the thermal state. 
Unfortunately, they are not straightforward to directly detect because of two factors:
(1) The soliton width in-situ, being a fraction of a $\mu$m, is significantly narrower than the usual detector resolution (typically several $\mu$m). However, a recently developed detection scheme based on scattering of an electron beam could overcome this difficulty \cite{SOtt}. Also, (2) an identification of solitons  by ``eyeballing'' is only straightforward if one has access to observations of the in-situ dynamics -- a single-time density slice typically shows many density dips, and it is not easy to distinguish the fast ``phonon-like'' ones from bona-fide solitons. 

\begin{figure}[hbt]
\includegraphics[width=8.6cm]{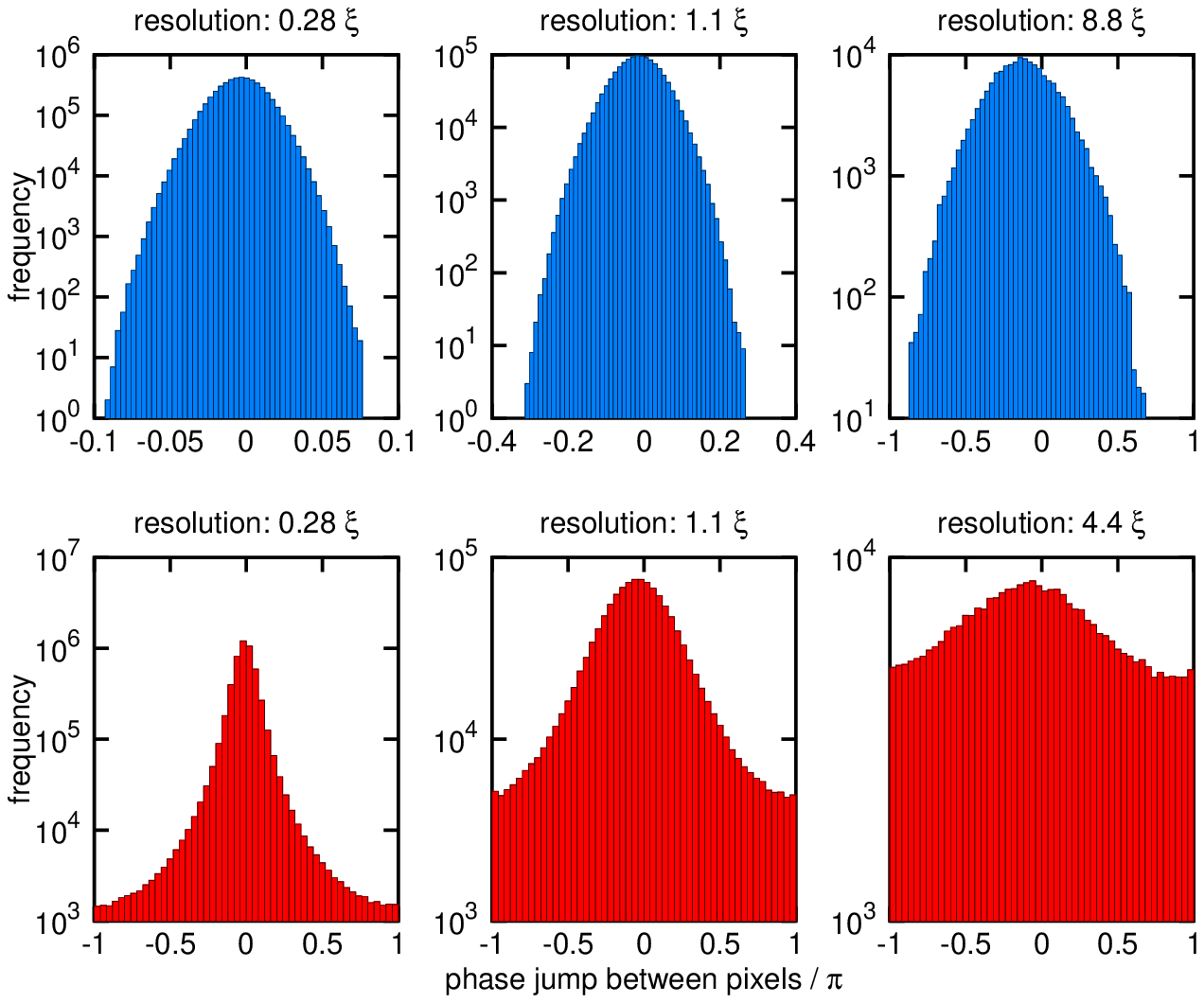}
\caption{Experimental signatures of thermal solitons in the uniform gas  --- phase-jump frequency (log scale) at $\gamma=0.02$ for low temperature (no deep solitons -- blue, $T=10^4$) and high temperature (many deep solitons -- red, $T=10\times 10^4$). Columns show the effect of different imaging resolutions in units of healing length $\xi$ ( $= 1/N\sqrt{\gamma}$ in our units), which is the half-width of a deep soliton. Resolution worsens from left (below $\xi$) to right (unable to resolve soliton dips). Note the strong resistance of the difference in shape of the distribution (parabolic / extended) even into regimes where individual solitons cannot be resolved. 
\label{f5}}
\end{figure}

However, several approaches hold promise of overcoming one or both of the above issues. One is a direct observation of deep solitons in an expanding cloud. After about $10$ms of expansion of a typical cloud, the soliton size can exceed detector resolution. The method was successfully used in engineered soliton experiments \cite{SSengstock, SPhillips, SSengstock1}. 

Another promising approach is to look for the expected large phase jump between phase domains that occurs at the soliton. If relatively recognizable phase domains are present between solitons, then the phase jump of close to $\pm\pi$ that occurs at deep soliton defects should be detectable with an imaging resolution that is sufficient only to resolve the phase domains \cite{SOberthaler}. These are much wider than the solitons if the soliton density is not extreme. 

It turns out that even with very noisy domains, a qualitative difference can be observed between clouds with and without solitons. Fig.~S\ref{f5} shows the phase-jump histogram (for phase jumps between neighboring pixel pairs) at $\gamma=0.02$ for both low temperature and high temperature. The low-temperature system has a Gaussian distribution of phase jumps (parabolic on the plot, which has a log scale), which broadens but does not change shape as the resolution is worsened. This reflects the addition of more and more random small phase fluctuations that arise primarily from Bogoliubov excitations. In stark contrast, when deep solitons are present, phase jumps of $\pi$ appear quite frequently, and -- most significantly -- the distribution flattens out for large jumps. This qualitative difference (flattening-out / Gaussian) survives even to resolutions that are incapable of resolving the actual density dip of the solitons. 

The anti-trapping technique may also be quite promising. This technique has been used to image short wavelength shock waves in elongated gases \cite{SChangEngels}, which have some phenomenological features in common with solitons. An inverted parabolic potential is rapidly applied to the gas along the long direction to prevent the evolution of in-situ density to momentum density in that direction, but to instead magnify the density profile. 

Finally, a recent work has presented signatures of solitons in the form of characteristic features of the momentum distribution\cite{SGasenzer}. It analyzed solitons formed after a quantum quench, but is also appropriate for the randomly placed thermal solitons discussed here.

\end{document}